\documentclass[11pt]{revtex4}
\usepackage[pdftex]{graphicx}
\usepackage{amsmath}
\usepackage{amsfonts}
\usepackage{amssymb}
\newcommand\spm{\mathrel{\text{\framebox[0.9\width]{$\pm$}}}}
\newcommand\smp{\mathrel{\text{\framebox[0.9\width]{$\mp$}}}}
\newcommand\cpm{\mathrel{\text{\textcircled{\makebox{$\pm$}}}}}
\newcommand\cmp{\mathrel{\text{\textcircled{\makebox{$\mp$}}}}}
\begin{document}
\title{Families of spinors in $d=(1+5)$ with zweibein and two kinds of spin 
connection fields on  an almost $S^2$} 
\author{D. Lukman and N.S. Manko\v c Bor\v stnik \\
Department of Physics, FMF, University of Ljubljana,\\
Jadranska 19, SI-1000 Ljubljana, Slovenia
}

\begin{abstract} 
We studied~\cite{hn,dn} properties of spinors in a toy model in $d=(1+5)$ as a 
step towards realistic Kaluza-Klein (like) theories in non compact spaces. 
${\cal M}^{(5+1)}$ was assumed to break to an infinite disc with a
zweibein which makes a disc curved on $S^2$ and with a spin connection field  
which allows on such a sphere only one massless spinor state.  
This time we are taking into account families of spinors interacting with 
several spin connection fields, as required for this toy model by the 
{\it spin-charge-family} theory~\cite{norma92939495,NF,NPLB}.  
We are studying possible masslesness of families of spinors: Spinors regroup  
into subgroups of an even number of families.  
\end{abstract}
\maketitle
\section{Introduction}
\label{introduction}

The {\it spin-charge-family} theory~\cite{norma92939495,NF}, proposed by N.S. Manko\v c 
Bor\v stnik,  is offering an explanation for the appearance of families of fermions in 
any dimension. 
Starting in $d=(13 + 1)$ with a simple action for massless fermions interacting with the 
gravitational interaction only -- that is with the vielbeins and two kinds of the spin 
connection fields -- the theory manifests effectively at low energies the so far observed 
properties of fermions and bosons, explaining the assumptions of the {\it standard model}, 
among them the appearance of families, the Higgs scalar and Yukawa couplings, and predicting
the fourth family, a stable fifth family (explaining the existence of the dark matter) and 
several scalar fields, which could be observed at the LHC. 

A simple toy model~\cite{hn,dn,dhn}, which includes families as in the {\it spin-charge-family} 
theory~\cite{norma92939495,NF}, 
is expected to help to better understand properties of families of spinors at observable energies.  
This contribution is the continuation of the ref.~\cite{dhn}, where families were already 
included, and so were two kinds of the spin connection fields. We make this time a small step 
further in understanding properties of families of spinors, allowing the influence of several  
spin connection fields. 
We start with a massless spinor in  a flat manifold ${\cal M}^{(5+1)}$, which breaks into 
${\cal M}^{(3+1)}$ times an infinite disc. The vielbein on the disc curves the disc into 
(almost) a sphere $S^{2}$
\begin{eqnarray}
e^{s}{}_{\sigma} = f^{-1}
\begin{pmatrix}1  & 0 \\
 0 & 1 
 \end{pmatrix},
f^{\sigma}{}_{s} = f
\begin{pmatrix}1 & 0 \\
0 & 1 \\
\end{pmatrix},
\label{fzwei}
\end{eqnarray}
with 
\begin{eqnarray}
\label{f}
f &=& 1+ (\frac{\rho}{2 \rho_0})^2= \frac{2}{1+\cos \vartheta},\nonumber\\ 
x^{(5)} &=& \rho \,\cos \phi,\quad  x^{(6)} = \rho \,\sin \phi, \quad E= f^{-2}.
\end{eqnarray}
The angle $\vartheta$ is the ordinary azimuthal angle on a sphere. 
The last relation follows  from $ds^2= 
e_{s \sigma}e^{s}{}_{\tau} dx^{\sigma} dx^{\tau}= f^{-2}(d\rho^{2} + \rho^2 d\phi^{2})$.
We use indices $s,t=5,6$ to describe the flat index in the space of an infinite plane, and 
$\sigma, \tau = (5), (6), $ to describe the Einstein index.  
Rotations around  the axis through the two poles of a sphere are described by the angle $\phi$, 
while $\rho = 2 \rho_0 \, \sqrt{\frac{1- \cos \vartheta}{1+ \cos \vartheta}}$. 
The volume of this non compact sphere is finite, equal to $V= \pi\, (2 \rho_0)^2$.  The symmetry 
of $S^2$ is a symmetry of $U(1)$ group. We  look for  chiral fermions on this sphere, that is 
the fermions of 
only one handedness and accordingly mass protected, without including any extra fundamental gauge 
fields to the action from Eq.(\ref{action}). We study  the influence of several spin connection 
fields on properties of families. 

We take into account that there are two kinds of the $\gamma$ operators, beside the Dirac 
$\gamma^a$ also $\tilde{\gamma}^a$ introduced in~\cite{norma92939495,NF,holgernorma20023}. 
Correspondingly the covariant momentum of spinor is 
\begin{eqnarray}
\label{covmom}
p_{0a}&=& f^{\alpha}{}_a\, p_{\alpha} + \frac{1}{2E}\,\{p_{\alpha}, f^{\alpha}{}_a E\}_- 
-\frac{1}{2} S^{cd}  \omega_{cda}- \frac{1}{2} \tilde{S}^{cd} \tilde{\omega}_{cda}\,, \nonumber\\
S^{ab} &=& \frac{i}{4} (\gamma^a \gamma^b- \gamma^b \gamma^a)\,, \quad\quad  
\tilde{S}^{ab}= \frac{i}{4} (\tilde{\gamma}^a \tilde{\gamma}^b- \tilde{\gamma}^b \tilde{ \gamma}^a)\,,
\end{eqnarray}
with $ E = \det(e^a{\!}_{\alpha}) $ and with vielbeins $f^{\alpha}{\!}_{a}$~\footnote{$f^{\alpha}{}_{a}$ 
are inverted vielbeins to $e^{a}{}_{\alpha}$ with the properties $e^a{}_{\alpha} f^{\alpha}{\!}_b = $ 
$\delta^a{}_b,\; e^a{}_{\alpha} f^{\beta}{}_a = $ $\delta^{\beta}_{\alpha} $. 
Latin indices  
$a,b,..,m,n,..,s,t,..$ denote a tangent space (a flat index),
while Greek indices $\alpha, \beta,..,\mu, \nu,.. \sigma,\tau ..$ denote an Einstein 
index (a curved index). Letters  from the beginning of both the alphabets
indicate a general index ($a,b,c,..$   and $\alpha, \beta, \gamma,.. $ ), 
from the middle of both the alphabets   
the observed dimensions $0,1,2,3$ ($m,n,..$ and $\mu,\nu,..$), indices from 
the bottom of the alphabets
indicate the compactified dimensions ($s,t,..$ and $\sigma,\tau,..$). 
We assume the signature $\eta^{ab} =
diag\{1,-1,-1,\ldots,-1\}$.
}, the gauge fields of the infinitesimal generators of translation, and  with the two kinds of the 
spin connection fields: i. $\omega_{ab\alpha}$,  the gauge fields of  $S^{ab}$ 
and ii. $\tilde{\omega}_{ab\alpha}$,  the gauge fields of  $\tilde{S}^{ab}$. 

We make a choice  of the spin connection fields of the two kinds on the infinite disc as follows 
(assuming that there must be some fermion sources causing these spin connections)
\begin{eqnarray}
  f^{\sigma}{}_{s'}\, \omega_{st \sigma} &=& i F_{56}\, f \, \varepsilon_{st}\; 
  \frac{e_{s' \sigma} x^{\sigma}}{(\rho_0)^2}\,= 
  -\frac{1}{2E}\{p_{\sigma}, Ef^{\sigma}{}_{s'}\}_{-} \; \varepsilon_{st}\, 4 F_{56}\,, 
  \nonumber\\
  f^{\sigma}{}_{s'}\, \tilde{\omega}_{st \sigma} &=& i \tilde{F}_{56}\, f \, \varepsilon_{st}\; 
  \frac{e_{s' \sigma} x^{\sigma}}{(\rho_0)^2}\,
  = -\frac{1}{2E}\{p_{\sigma}, Ef^{\sigma}{}_{s'}\}_{-} \;\varepsilon_{st}\, 4 \tilde{F}_{56}\,, \nonumber\\
  f^{\sigma}{}_{s}\, \tilde{\omega}_{mn \sigma} &=&  -\frac{1}{2E}\{p_{\sigma}\,,  
  E f^{\sigma}{}_{s}\}_{-} \; 4 \tilde{F}_{mn}\;, \tilde{F}_{mn}= - \tilde{F}_{nm}\,,\nonumber\\
  &&\quad s=5,6,\,\,\; \sigma=(5),(6)\,. 
\label{omegas}
\end{eqnarray}
We take the starting action in agreement with the  {\it spin-charge-family} theory
for this toy model in  $d=(5+1)$, that is the action for a massless spinor  (${\cal S}_{f}$) with the 
covariant momentum $p_{0a}$ from Eq.~(\ref{covmom}) interacting with gravity only and for the 
vielbein and the two kinds of the spin connection  fields (${\cal S}_{b}$) 
  \begin{eqnarray}
         {\cal S} &=& {\cal S}_{b} +  {\cal S}_{f}\,,\nonumber\\
  {\cal S}_{b} =  \int \; d^d x \, E \, (\alpha {\cal\,  R} + \tilde{\alpha} \tilde{{\cal\,  R}})\,, 
  &\quad &\,\;   {\cal L}_{f}= \psi^{\dagger}\, \gamma^0 \gamma^a \,E\,  p_{0a}\,\psi
  \,.
  \label{action}
  \end{eqnarray}
The two Riemann scalars,  ${\cal R} = {\cal R}_{abcd}\,\eta^{ac}\eta^{bd}$ and 
$\tilde{{\cal R}} = \tilde{{\cal R}}_{abcd}\,\eta^{ac}\eta^{bd}$, are
determined by the Riemann tensors
%
$     {\cal R}_{abcd}      = \frac{1}{2}\, f^{\alpha}{}_{[a} f^{\beta}{}_{b]}(\omega_{cd \beta, \alpha} 
- \omega_{ce \alpha} \omega^{e}{}_{d \beta} )\;$, 
$\tilde{{\cal R}}_{abcd}  =  \frac{1}{2}\, f^{\alpha [ a} f^{\beta b ]} \; (\tilde{\omega}_{cd \beta,\alpha} - 
\tilde{\omega}_{c e \alpha} \tilde{\omega}^{e}{}_{d \beta})\; $\,, $[a\,\,b]$ means that the 
anti-symmetrization must be performed over the two indices $a$ and $b$.

We assume no gravity in $d=(3+1)$: $f^{\mu}{}_m = \delta^{\mu}_m$ and $\omega_{mn\mu}=0$ for 
$ m,n=(0,1,2,3), \;  \mu =(0,1,2,3)$. Accordingly $(a,b,\dots)$ run in Eq.~(\ref{action}) only over 
$s \in (5,6)$. 
Taking into account the subgroup structure of the operators  $\tilde{S}^{mn}$ 
\begin{eqnarray}
\label{so31sub}
\vec{\tilde{N}}^{\cpm}&=&\frac{1}{2} (\tilde{S}^{23}\pm i \tilde{S}^{01},\tilde{S}^{31}\pm i \tilde{S}^{02}, 
\tilde{S}^{12}\pm i \tilde{S}^{03} )\,, \nonumber\\
\tilde{N}^{\oplus \spm}&=&\tilde{N}^{\oplus 1}  \pm i\, \tilde{N}^{\oplus 2}\,, \quad 
\tilde{N}^{\ominus \spm} = \tilde{N}^{\ominus 1} \pm i\, \tilde{N}^{\ominus 2}\,,
\end{eqnarray}
we can rewrite the $\frac{1}{2} \tilde{S}^{cd} \tilde{\omega}_{cda}$ part of the covariant momentum 
(Eq.~\ref{covmom}) as follows 
\begin{eqnarray}
\label{omegamnFmn}
-\frac{1}{2}\, f \,  \tilde{S}^{mn} \tilde{\omega}_{mn\pm}  &=&
     \sum_i \tilde{N}^{\oplus i} \tilde{A}^{\oplus i}_{\pm} + \sum_i
     \tilde{N}^{\ominus i} \tilde{A}^{\ominus i}_{\pm} \nonumber\\
     &=&  \tilde{N}^{\oplus \boxplus } \tilde{A}^{\oplus \boxplus}_{\pm} + 
     \tilde{N}^{\oplus \boxminus } \tilde{A}^{\oplus \boxminus}_{\pm} +  
     \tilde{N}^{\oplus 3 } \tilde{A}^{\oplus 3}_{\pm}+ 
  \tilde{N}^{\ominus \boxplus } \tilde{A}^{\ominus \boxplus}_{\pm} + 
     \tilde{N}^{\ominus \boxminus } \tilde{A}^{\ominus \boxminus}_{\pm} +  
     \tilde{N}^{\ominus 3 } \tilde{A}^{\ominus 3}_{\pm
     } \,,\nonumber\\
     \tilde{\omega}_{mn\pm} &=& \tilde{\omega}_{mn 5} \mp i \tilde{\omega}_{mn 6}\,. 
\end{eqnarray}
The notation was used
\begin{eqnarray}
\label{so13}
f^{\sigma}{}_{s} \,\tilde{A}^{\cpm i}_{\sigma} &=&- f^{\sigma}{}_{s} \{
(\tilde{\omega}_{23 \sigma} \mp i \tilde{\omega}_{01 \sigma}), 
(\tilde{\omega}_{31 \sigma} \mp i \tilde{\omega}_{02 \sigma}),
(\tilde{\omega}_{12 \sigma} \mp i \tilde{\omega}_{03 \sigma})\}\,\nonumber\\
&=&  \, \delta^{\sigma}_{s }\; \frac{1}{2E}\{p_{\sigma}, Ef \}_{-} \; 4 \left ( 
\tilde{F}^{\cpm 1}, \tilde{F}^{\cpm 2}, \tilde{F}^{\cpm 3} \right)\,,\nonumber\\
\tilde{A}^{\oplus \spm}_{s} &=&  \frac{1}{2} \,(\tilde{A}^{\oplus 1}_{s}
\mp i \, \tilde{A}^{\oplus 2}_{s})\,,\quad 
\tilde{A}^{\ominus \spm}_{s} = \frac{1}{2} \,(\tilde{A}^{\ominus 1}_{s}
\mp i \, \tilde{A}^{\ominus 2}_{s})\,,\nonumber\\
\tilde{F}^{\oplus \spm}&=& (\tilde{F}^{23}\mp \tilde{F}^{02}) - i (\pm \tilde{F}^{31}+\tilde{F}^{01})\,,\quad
\tilde{F}^{\oplus 3} = (\tilde{F}^{12}- i \tilde{F}^{03})\,, \nonumber\\
\tilde{F}^{\ominus \spm}&=& (\tilde{F}^{23}\pm \tilde{F}^{02}) + i (\mp \tilde{F}^{31}+\tilde{F}^{01})\,,\quad
\tilde{F}^{\ominus 3} = (\tilde{F}^{12}+ i \tilde{F}^{03})\,, \nonumber\\
\sigma &=& ((5),(6))\,,\quad s= (5,6)\,, 
\end{eqnarray}
with $\omega_{abc}$ and $\tilde{\omega}_{abc}$  defined in Eq.~(\ref{omegas}).

We  study  intervals within   which the parameters  of both kinds of the spin connection fields 
($F_{56}$, $\tilde{F}_{56}$, $\tilde{F}^{\oplus \spm} $,  $\tilde{F}^{\oplus 3}$, $\tilde{F}^{\ominus \spm} $,  
$\tilde{F}^{\ominus 3}$) allow massless solutions of the equation 
\begin{eqnarray}
&&\{E\gamma^0 \gamma^m p_m + E f \gamma^0 \gamma^s \delta^{\sigma}_s  ( p_{0\sigma} 
+  \frac{1}{2 E f}
\{p_{\sigma}, E f\}_- )\}\psi=0,\quad {\rm with} \nonumber\\
&& p_{0\sigma} = p_{\sigma}- \frac{1}{2} S^{st}\omega_{st \sigma} - \frac{1}{2} \tilde{S}^{ab} 
\tilde{\omega}_{ab \sigma}\,, 
\label{weylE}
\end{eqnarray}
for one or several families of spinors. 
To solve Eq.~(\ref{weylE}) we must tell more about the appearance of families of spinors.

\section{Solutions of equations of motion  for families of spinors}
\label{equations}

We first briefly explain, following the refs.~\cite{NF,NPLB, hn,dn,dhn}, the appearance of families 
in our toy model, using what is called the technique~\cite{holgernorma20023}. 
There are $2^{d/2 -1}=4$ families in our toy model, each family with $2^{d/2 -1}=4$ members. 
In the technique~\cite{holgernorma20023} the states are defined as a product of nilpotents 
and projectors  
%
\begin{eqnarray}
\stackrel{ab}{(\pm i)}: &=& \frac{1}{2}(\gamma^a \mp  \gamma^b),  \; 
\stackrel{ab}{[\pm i]}: = \frac{1}{2}(1 \pm \gamma^a \gamma^b), \quad
{\rm for} \,\; \eta^{aa} \eta^{bb} = -1, \nonumber\\
\stackrel{ab}{(\pm )}: &= &\frac{1}{2}(\gamma^a \pm i \gamma^b),  \; 
\stackrel{ab}{[\pm ]}: = \frac{1}{2}(1 \pm i\gamma^a \gamma^b), \quad
{\rm for} \,\; \eta^{aa} \eta^{bb} =1,
\label{snmb:eigensab}
\end{eqnarray} 
which are the eigen vectors  of $S^{ab}$ as well as of $\tilde{S}^{ab}$ as follows
\begin{eqnarray}
S^{ab} \stackrel{ab}{(k)} =  \frac{k}{2} \stackrel{ab}{(k)}, \quad 
S^{ab} \stackrel{ab}{[k]} =  \frac{k}{2} \stackrel{ab}{[k]}, \quad
\tilde{S}^{ab} \stackrel{ab}{(k)}  = \frac{k}{2} \stackrel{ab}{(k)},  \quad 
\tilde{S}^{ab} \stackrel{ab}{[k]}  =   - \frac{k}{2} \stackrel{ab}{[k]}\;,
\label{snmb:eigensabev}
\end{eqnarray}
with the properties that $\gamma^a$ transform   
$\stackrel{ab}{(k)}$ into  $\stackrel{ab}{[-k]}$, while 
$\tilde{\gamma}^a$ transform  $\stackrel{ab}{(k)}$ 
into $\stackrel{ab}{[k]}$ 
\begin{eqnarray}
\gamma^a \stackrel{ab}{(k)}= \eta^{aa}\stackrel{ab}{[-k]},\; 
\gamma^b \stackrel{ab}{(k)}= -ik \stackrel{ab}{[-k]}, \; 
\gamma^a \stackrel{ab}{[k]}= \stackrel{ab}{(-k)},\; 
\gamma^b \stackrel{ab}{[k]}= -ik \eta^{aa} \stackrel{ab}{(-k)}\,,\nonumber\\
\label{snmbgraph}
\tilde{\gamma^a} \stackrel{ab}{(k)} = - i\eta^{aa}\stackrel{ab}{[k]},\;
\tilde{\gamma^b} \stackrel{ab}{(k)} =  - k \stackrel{ab}{[k]}, \;
\tilde{\gamma^a} \stackrel{ab}{[k]} =  \;\;i\stackrel{ab}{(k)},\; 
\tilde{\gamma^b} \stackrel{ab}{[k]} =  -k \eta^{aa} \stackrel{ab}{(k)}\,. 
\end{eqnarray}
After making a choice of the Cartan subalgebra, for which we take: ($S^{03}, S^{12}, S^{56}$) 
and ($\tilde{S}^{03}, \tilde{S}^{12}, \tilde{S}^{56}$), the four spinor families, each with 
four vectors, which are  eigen vectors of the chosen Cartan subalgebra with the eigen values 
from Eq.~(\ref{snmb:eigensabev}), can be found in~\cite{dhn} written as four times four products of 
projectors $\stackrel{ab}{[k]}$ and nilpotents $\stackrel{ab}{(k)}$. We present here only one 
member of each family, the one with  eigen values of $S^{56}$ and $S^{12}$ both equal to $\frac{1}{2}$, 
and of left handedness with respect to $d=(5+1)$ and $d=(3+1)$ 
\begin{align}
\label{weylrep}
\varphi^{1 I}_{1} &= \stackrel{56}{(+)} \stackrel{03}{(+i)} \stackrel{12}{(+)}\psi_0,
&\varphi^{1 II}_{1} &= \stackrel{56}{(+)} \stackrel{03}{[+i]} \stackrel{12}{[+]}\psi_0, \nonumber\\
&&
\nonumber\\
\varphi^{1 III}_{1} &= \stackrel{56}{[+]} \stackrel{03}{[+i]}\stackrel{12}{(+)}\psi_0,
&\varphi^{1 IV}_{1} &= \stackrel{56}{[+]} \stackrel{03}{(+i)} \stackrel{12}{[+]}\psi_0,\nonumber\\
\end{align}
where  $\psi_0$ is a vacuum  for the spinor state. One can reach  from the first member  
$\varphi^{1 I}_{1}$ of the first family the same  family member of  all the other families by 
the application of $\tilde{S}^{ab}$.
The reader can create the remaining three members of each family  
by applying the generators $S^{ab}$ on the presented member~\cite{dhn}. The  rest of the members of, say,  
the first family  are: $\varphi^{1 I}_{2} =\stackrel{56}{(+)}  \stackrel{03}{[-i]} 
\stackrel{12}{[-]}\psi_0\,,$    $\varphi^{2 I}_{1} =\stackrel{56}{[-]}  \stackrel{03}{[-i]} 
\stackrel{12}{(+)}\psi_0\,,$    $\varphi^{2 I}_{2} =\stackrel{56}{[-]} \stackrel{03}{(+i)} 
\stackrel{12}{[-]}\psi_0\,$. 
If we write the operators of handedness in $d=(1+5)$ as $\Gamma^{(1+5)} = \gamma^0 \gamma^1 
\gamma^2 \gamma^3 \gamma^5 \gamma^6$ ($= 2^3 i S^{03} S^{12} S^{56}$), in $d=(1+3)$ 
as $\Gamma^{(1+3)}= -i\gamma^0\gamma^1\gamma^2\gamma^3 $ ($= 2^2 i S^{03} S^{12}$) 
and in the two dimensional space as $\Gamma^{(2)} = i\gamma^5 \gamma^6$ 
($= 2 S^{56}$), we find that all the  states of all the families are left handed with respect to 
$\Gamma^{(1+5)}$, with the eigen value $-1$, the first two states of the first family, and 
correspondingly the first two states of any family, are right handed and the second two 
 states are left handed with respect to $\Gamma^{(2)}$, with  the eigen values $1$ and $-1$, 
 respectively, while the first two are left handed 
and the second two right handed with respect to $\Gamma^{(1+3)}$ with the eigen values $-1$ and $1$, 
respectively. 
Having the rotational symmetry around the axis perpendicular to the plane of the fifth and the sixth 
dimension we require that $\psi^{(6)}$ is the eigen function of the total angular momentum
operator $M^{56}= x^5 p^6-x^6 p^5  + S^{56}= -i \frac{\partial}{\partial \phi} + S^{56}$
\begin{eqnarray}
M^{56}\psi^{(6)}= (n+\frac{1}{2})\,\psi^{(6)}\,.
\label{mabx}
\end{eqnarray}

Accordingly we write, when taking into account Eq.~(\ref{weylrep}),  
the most general wave function  
$\psi^{(6)}$ obeying Eq.~(\ref{weylE}) in $d=(1+5)$ as
\begin{eqnarray}
\psi^{(6)}= {\cal N}\, \sum_{i=I,II,III,IV}\, ({\cal A}^{i}_{n}\, {\stackrel{56}{(+)}}{}^{i}\,
\psi^{(4i)}_{(+)}  
+ {\cal B}^{i}_{n+1}\, e^{i \phi}\, {\stackrel{56}{[-]}}{}^{i}\, \psi^{(4 i)}_{(-)})\, e^{in \phi}.
\label{mabpsi}
\end{eqnarray}
where ${\cal A}^{i}_{n}$ and ${\cal B}^{i}_{n}$ depend on $x^{\sigma}$, while $\psi^{(4 i)}_{(+)}$ 
and $\psi^{(4 i)}_{(-)}$  determine the spin 
and the coordinate dependent parts of the wave function $\psi^{(6)}$ in $d=(3+1)$ in accordance 
with the definition in Eq.(\ref{weylrep}), for example, 
\begin{eqnarray}
\psi^{(4 I)}_{(+)} &=& \alpha^{I}_+ \; {\stackrel{03}{(+i)}}\, {\stackrel{12}{(+)}} + 
\beta^{I}_+ \; {\stackrel{03}{[-i]}}\, {\stackrel{12}{[-]}}, \nonumber\\ 
\psi^{(4 I)}_{(-)}&=& \alpha^{I}_- \; {\stackrel{03}{[-i]}}\, {\stackrel{12}{(+)}} + 
\beta^{I}_- \; {\stackrel{03}{(+i)}}\, {\stackrel{12}{[-]}}. 
\label{psi4}
\end{eqnarray}
${\stackrel{56}{(+)}}{}^{i}= {\stackrel{56}{(+)}},$ for $i=I,II$ and ${\stackrel{56}{(+)}}{}^{i}=
{\stackrel{56}{[+]}}$ for $i=III,IV$, while ${\stackrel{56}{[-]}}{}^{i}= 
{\stackrel{56}{[-]}}$ for $i=I,II$ and ${\stackrel{56}{[-]}}{}^{i}= {\stackrel{56}{(-)}}$ for $i=III,IV$.
Using $\psi^{(6)}$ in Eq.~(\ref{weylE}) and separating dynamics in $(1+3)$ and on $S^2$, 
the following relations follow, from which we recognize the mass term $m^{I}$:  
$\frac{\alpha^{i}_{+}}{\alpha^{i}_-} (p^0-p^3) - \frac{\beta^{i}_+}{\alpha^{i}_-} (p^1-ip^2)= m^{i},$ 
$\frac{\beta^{i}_+}{\beta^{i}_-} (p^0+p^3) - \frac{\alpha^{i}_+}{\beta^{i}_-} (p^1+ip^2)= m^{i},$ 
$\frac{\alpha^{i}_-}{\alpha^{i}_+} (p^0+p^3) + \frac{\beta_-}{\alpha_+} (p^1-ip^2)= m^{i},$
$\frac{\beta^{i}_-}{\beta^{i}_+} (p^0-p^3) + \frac{\alpha^{i}_-}{\beta^{i}_+} (p^1-ip^2)= m^{i}.$ 
(One notices that for massless solutions  ($m^{i}=0$)  $\psi^{(4i)}_{(+)}$ 
and $\psi^{(4i)}_{(-)}$, for each $i=  I,II,III,IV,$ 
decouple.) 

For a spinor with the momentum $p^m= (p^0,0,0,p^3)$ in $d=(3+1)$ the spin and coordinate dependent parts 
for four families is:  $\psi^{(4 I)}_{(+)} = $ 
$\alpha \; {\stackrel{03}{(+i)}}\, {\stackrel{12}{(+)}}\,$, $\psi^{(4 II)}_{(+)} = $ 
$\alpha\; {\stackrel{03}{[+i]}}\, {\stackrel{12}{[+]}}\,$, $\psi^{(4 III)}_{(+)} = $ 
$\alpha\; {\stackrel{03}{[+i]}}\, {\stackrel{12}{(+)}}\,$, $\psi^{(4 IV)}_{(+)} = $ 
$\alpha \; {\stackrel{03}{(+i)}}\, {\stackrel{12}{[+]}}\,$.

Taking the above derivation into account (Eqs.~(\ref{mabpsi}, \ref{f}, \ref{omegas}, \ref{psi4},
\ref{so31sub}, \ref{omegamnFmn}, \ref{so13}))  the equation of motion for spinors follows~\cite{dhn} 
from the action~(\ref{action})
 \begin{eqnarray}
 \label{weylErho}
 if \,&& \{ e^{i \phi 2S^{56}}\, [(\frac{\partial}{\partial \rho} + \frac{i\, 2 S^{56}}{\rho} \, 
 (\frac{\partial}{\partial \phi}) ) -  \frac{1}{2 \,f} \, \frac{\partial f}{\partial \rho }\, 
 (1- 2 F_{56} \, 2S^{56} - 2 \tilde{F}_{56}\, 2\tilde{S}^{56}  \nonumber\\
 && -  2 \tilde{F}^{\ominus\,\boxplus}   \, 2 \tilde{N}^{\ominus\,\boxplus} -
       2 \tilde{F}^{\ominus\,\boxminus}  \, 2 \tilde{N}^{\ominus\,\boxminus}-
       2 \tilde{F}^{\ominus\,3}          \, 2 \tilde{N}^{\ominus\,3} 
       \nonumber\\
 && -  2 \tilde{F}^{\oplus\,\boxplus}    \, 2 \tilde{N}^{\oplus\,\boxplus} -
       2 \tilde{F}^{\oplus\,\boxminus}   \, 2 \tilde{N}^{\oplus\,\boxminus}-
       2 \tilde{F}^{\oplus\,3}           \, 2 \tilde{N}^{\oplus\,3}  )\,]
 \, \}  \psi^{(6)}\nonumber\\
 + && \gamma^0 \gamma^5 \, m \, \psi^{(6)}=0.
 \end{eqnarray}
One easily recognizes that, due to the break of ${\cal M}^{(5+1)}$ into ${\cal M}^{(3+1)} \times$ 
an infinite disc, which concerns  (by our assumption) $S^{ab}$ and $\tilde{S}^{ab}$ sector, there 
are two times two coupled families: The first and the second, and the third and the fourth, while 
the first and the second  remain decoupled from the third and the fourth.
We end up with  two decoupled groups of equations of motion~\cite{dhn} 
(Eqs.~(16.26-16.33)) (which all depend on the parameters $F_{56}$ and $\tilde{F}_{56}$):\\
i. The equations for the first and the second family
\begin{eqnarray}
\label{mass12}
&&-if \bigl\{ [(\frac{\partial}{\partial\rho} - \frac{n}{\rho}) 
           - \frac{1}{2f} \frac{\partial f}{\partial\rho} 
           (1-2F_{56}-2\tilde{F}_{56}-2\tilde{F}^{\ominus 3})]\,  \mathcal{A}_n^I \nonumber\\
           &-& \frac{1}{2f} \frac{\partial f}{\partial\rho}
           \, 2\tilde{F}^{\ominus \boxplus}\, \mathcal{A}_n^{II}
    \bigr\} 
    + m \, \mathcal{B}_{n+1}^I = 0\,,\nonumber
\end{eqnarray}
\begin{eqnarray}
\label{mass12}
&&-if \bigl\{ [(\frac{\partial}{\partial\rho} + \frac{n+1}{\rho}) 
           - \frac{1}{2f} \frac{\partial f}{\partial\rho} 
                (1+2F_{56}-2\tilde{F}_{56} - 2\tilde{F}^{\ominus 3})] \, \mathcal{B}_{n+1}^{I} \nonumber\\
           &-& \frac{1}{2f} \frac{\partial f}{\partial\rho} 
           \, 2\tilde{F}^{\ominus \boxplus}\,  \mathcal{B}_{n+1}^{II}
   \bigr\} + m \, \mathcal{A}_{n}^{I} = 0\,,
\end{eqnarray}
\begin{eqnarray}
&&-if \bigl\{ [(\frac{\partial}{\partial\rho} - \frac{n}{\rho}) 
           - \frac{1}{2f} \frac{\partial f}{\partial\rho} 
            (1-2F_{56}-2\tilde{F}_{56}+ 2\tilde{F}^{\ominus 3})] \, \mathcal{A}_n^{II} \nonumber\\
           &-& \frac{1}{2f} \frac{\partial f}{\partial\rho}
           \, 2\tilde{F}^{\ominus \boxminus}\, \mathcal{A}_n^{I}
    \bigr\} + m \,\mathcal{B}_{n+1}^{II} = 0\,,\nonumber
    \end{eqnarray}
    \begin{eqnarray}
\label{mass12}
&&-if \bigl\{ [(\frac{\partial}{\partial\rho} + \frac{n+1}{\rho}) 
           - \frac{1}{2f} \frac{\partial f}{\partial\rho} 
              (1+2F_{56}-2\tilde{F}_{56}+2\tilde{F}^{\ominus 3})] \, \mathcal{B}_{n+1}^{II} \nonumber\\
           &-& \frac{1}{2f} \frac{\partial f}{\partial\rho} 
              \, 2\tilde{F}^{\ominus \boxminus}\, \mathcal{B}_{n+1}^{I}
    \bigl\} + m \,\mathcal{A}_{n}^{II} = 0\,.\nonumber
\end{eqnarray}
ii. The equations for the third and the fourth family 
\begin{eqnarray}
\label{mass34}
&&-if \bigl\{ [(\frac{\partial}{\partial\rho} - \frac{n}{\rho}) 
         - \frac{1}{2f} \frac{\partial f}{\partial\rho}
           (1-2F_{56}+2\tilde{F}_{56} -2\tilde{F}^{\oplus 3})] \, \mathcal{A}_n^{III}\nonumber\\
         &-& \frac{1}{2f} \frac{\partial f}{\partial\rho}
          \, (-2\tilde{F}^{\oplus \boxplus})\, \mathcal{A}_n^{IV}
    \bigr\} + m \,\mathcal{B}_{n+1}^{III} = 0\,,\nonumber
    \end{eqnarray}
    \begin{eqnarray}
\label{mass34}
&&-if \bigl\{ [(\frac{\partial}{\partial\rho} + \frac{n+1}{\rho}) 
          - \frac{1}{2f} \frac{\partial f}{\partial\rho} 
               (1+2F_{56}+2\tilde{F}_{56}-2\tilde{F}^{\oplus 3})] \, \mathcal{B}_{n+1}^{III}\nonumber\\
           &-& \frac{1}{2f} \frac{\partial f}{\partial\rho} 
             \,(-2\tilde{F}^{\oplus \boxplus })\, \mathcal{B}_{n+1}^{IV}
    \bigl\} + m \, \mathcal{A}_{n}^{III} = 0\,,
    \end{eqnarray}
\begin{eqnarray}
&&-if \bigl\{ [(\frac{\partial}{\partial\rho} - \frac{n}{\rho}) 
           - \frac{1}{2f} \frac{\partial f}{\partial\rho} 
           (1-2F_{56}+2\tilde{F}_{56}+2\tilde{F}^{\oplus 3})] \, \mathcal{A}_n^{IV}\nonumber\\
           &-& \frac{1}{2f} \frac{\partial f}{\partial\rho} 
             \,(-2\tilde{F}^{\oplus \boxminus })\, \mathcal{A}_n^{III}
    \bigr\} + m \,\mathcal{B}_{n+1}^{IV} = 0\,,\nonumber
    \end{eqnarray}
    \begin{eqnarray}
\label{mass34}
&&-if \bigl\{ [(\frac{\partial}{\partial\rho} + \frac{n+1}{\rho}) 
           - \frac{1}{2f} \frac{\partial f}{\partial\rho} 
                  (1+2F_{56}+2\tilde{F}_{56}+2\tilde{F}^{\oplus 3})] \, \mathcal{B}_{n+1}^{IV}\nonumber\\
           &-& \frac{1}{2f} \frac{\partial f}{\partial\rho} 
            \,(-2\tilde{F}^{\oplus \boxminus})\, \mathcal{B}_{n+1}^{III}
    \bigr\} + m \,\mathcal{A}_{n}^{IV} = 0\,. \nonumber
\end{eqnarray}
Let us look for possible normalizable~\cite{hn,dn} massless solutions for each of the two groups 
in dependence on the parameters which determine the strength of the spin connection fields. Both 
groups, although depending on different parameters of the spin connection fields, can be treated in 
an equivalent way. Let us therefore study massless solutions of the first group of equations of motion.

For $m=0$   the equations for $\mathcal{A}_n^{I}$ and $\mathcal{A}_n^{II}$ in Eq.~(\ref{mass12}) 
decouple from those for $\mathcal{B}_{n+1}^{I}$ and $\mathcal{B}_{n+1}^{II}$.
We get for massless solutions~\cite{anzemitjanorma}
\begin{eqnarray}
\label{group12Amassless}
\mathcal{A}_{n}^{I\pm }\;\,&=& a_{\pm}\;
\rho^{n}\, f^{\frac{1}{2}(1- 2F_{56} -2\tilde{F}_{56})}\;
f^{ \pm \sqrt{(\tilde{F}^{\ominus 3})^2+\tilde{F}^{\ominus \boxplus} \tilde{F}^{\ominus \boxminus}}} \,,\nonumber\\ 
\mathcal{A}_{n}^{II\pm}\;&=&  
\frac{\pm \sqrt{(\tilde{F}^{\ominus 3})^2+\tilde{F}^{\ominus \boxplus} \tilde{F}^{\ominus \boxminus}}
- \tilde{F}^{\ominus 3}}{\tilde{F}^{\ominus \boxplus}}\;\mathcal{A}_{n}^{I\pm }\,,
\nonumber\\
\mathcal{B}_{n+1}^{I \pm}&=& b_{\pm}\,
\rho^{-n-1}\, f^{\frac{1}{2}(1 + 2F_{56} -2\tilde{F}_{56})}\,
f^{\pm \sqrt{(\tilde{F}^{\ominus 3})^2+
\tilde{F}^{\ominus \boxplus} \tilde{F}^{\ominus \boxminus}}} \,,\nonumber\\ 
\mathcal{B}_{n+1}^{II\pm}&=& 
\frac{\pm \sqrt{(\tilde{F}^{\ominus 3})^2+\tilde{F}^{\ominus \boxplus} \tilde{F}^{\ominus \boxminus}}
- \tilde{F}^{\ominus 3}}{\tilde{F}^{\ominus \boxplus}}\;\mathcal{B}_{n+1}^{I\pm }\,,
\end{eqnarray}
$n$ is a positive integer.  The solutions ($\mathcal{A}_{n}^{I + }$, $\mathcal{A}_{n}^{II + }$) 
and ($\mathcal{A}_{n}^{I - }$, $\mathcal{A}_{n}^{II - }$) are two independent solutions, a general solution
is any superposition of these two. Similarly is true for 
($\mathcal{B}_{n+1}^{I \pm}$, $\mathcal{B}_{n+1}^{II \pm}$). 

In the massless case also $\mathcal{A}_{n}^{I,II\pm} $ 
decouple from $\mathcal{B}_{n+1}^{I,II\pm} $.

One can easily write down massless solutions of the second group of two families, 
decoupled from the first one, when knowing massless solutions of the first group of families. It follows
\begin{eqnarray}
\label{group12Amassless}
\mathcal{A}_{n}^{III \pm}\;\,&=& a_{\pm}\;
\rho^{n}\, f^{\frac{1}{2}(1- 2F_{56} + 2\tilde{F}_{56})}\;
f^{ \pm \sqrt{(\tilde{F}^{\oplus 3})^2+\tilde{F}^{\oplus \boxplus} \tilde{F}^{\oplus \boxminus}}} \,,\nonumber\\ 
\mathcal{A}_{n}^{IV \pm}\;&=&  
\frac{\pm \sqrt{(\tilde{F}^{\oplus 3})^2+\tilde{F}^{\oplus \boxplus} \tilde{F}^{\oplus \boxminus}}
- \tilde{F}^{\oplus 3}}{\tilde{F}^{\oplus \boxplus}}\;\mathcal{A}_{n}^{III \pm}\,,
\nonumber\\
\mathcal{B}_{n+1}^{III \pm}&=& b_{\pm}\,
\rho^{-n-1}\, f^{\frac{1}{2}(1 + 2F_{56} +2\tilde{F}_{56})}\,
f^{\pm \sqrt{(\tilde{F}^{\oplus 3})^2+
\tilde{F}^{\oplus \boxplus} \tilde{F}^{\oplus \boxminus}}} \,,\nonumber\\ 
\mathcal{B}_{n+1}^{IV \pm}&=& 
\frac{\pm \sqrt{(\tilde{F}^{\oplus 3})^2+\tilde{F}^{\oplus \boxplus} \tilde{F}^{\oplus \boxminus}}
- \tilde{F}^{\oplus 3}}{\tilde{F}^{\oplus \boxplus}}\;\mathcal{B}_{n+1}^{III\pm }\,,
\end{eqnarray}
$n$ is a positive integer.

Requiring that only normalizable (square integrable) solutions are acceptable 
\begin{eqnarray}
2\pi \, \int^{\infty}_{0} \,E\, \rho d\rho \,({\cal A}^{i \star}_{n} {\cal A}^{i}_{n}
+{\cal B}^{i\star}_{n} {\cal B}^{i}_{n} )&& < \infty\,, 
\label{masslesseqsolf}
\end{eqnarray}
$i\in\{I,II,III,IV\}\,$, one finds that $\mathcal{A}^{i}_{n}$ and $\mathcal{B}^{i}_{n}$ are 
normalizable~\cite{hn,dn,anzemitjanorma} under the following conditions 
\begin{eqnarray}
\label{norm-solAB1234}
& & \mathcal{A}^{I,II}_{n}\;\;\;\;\,: \;\,-1 < n < 2 (F_{56}+ \tilde{F}_{56} 
\pm \sqrt{(\tilde{F}^{\ominus 3})^2+ \tilde{F}^{\ominus \boxplus} \tilde{F}^{\ominus \boxminus} }\,)\,, 
\nonumber\\
& &\mathcal{B}^{I,II}_{n}\;\;\;\;\,:\;\, 2(F_{56} - \tilde{F}_{56}\pm  \sqrt{(\tilde{F}^{\ominus 3})^2+
 \tilde{F}^{\ominus \boxplus} \tilde{F}^{\ominus \boxminus} })< n <1 \, ,
 \nonumber\\
& &\mathcal{A}^{III,IV}_{n}\,: \;\,-1 < n < 2 (F_{56}- \tilde{F}_{56} 
\pm \sqrt{(\tilde{F}^{\oplus 3})^2+ \tilde{F}^{\oplus \boxplus} \tilde{F}^{\oplus \boxminus} }\,)\,, 
\nonumber\\
& &\mathcal{B}^{III,IV}_{n}\,:\;\, 2(F_{56} + \tilde{F}_{56}\pm  \sqrt{(\tilde{F}^{\oplus 3})^2+
 \tilde{F}^{\oplus \boxplus} \tilde{F}^{\oplus \boxminus} })< n <1 \, .
\end{eqnarray}
One immediately sees that for $F_{56}=0=\tilde{F}_{56}$ there is no solution for the zweibein from Eq.~(\ref{f}). 
Let us first assume that $\tilde{F}^{\cmp i}=0\,; \,i\in\{1,2,3\}$.  
Eq.~(\ref{norm-solAB1234}) tells us that the strengths $F_{56}, \tilde{F}_{56}$ of the spin connection fields 
($\omega_{56 \sigma}$ and $\tilde{\omega}_{56 \sigma}$) can make a choice between the 
massless solutions (${\cal A}^{I,II}_n$, ${\cal A}^{III,IV}_n$) and (${\cal B}^{I,II}_n,  {\cal B}^{III,IV}_n$):\\ 
For 
\begin{eqnarray}
0< 2(F_{56}+ \tilde{F}_{56}) \le 1,\quad \tilde{F}_{56} < F_{56}
\label{Fformassless}
\end{eqnarray}
there exist four massless left handed solutions  with respect 
to $(1+3)$. For  
\begin{eqnarray}
0< 2(F_{56}+ \tilde{F}_{56}) \le 1,\quad \tilde{F}_{56} = F_{56}
\label{Fformassless2}
\end{eqnarray}
 the only massless solution are the two left handed spinors with respect to 
$(1+3)$
\begin{eqnarray}
\psi^{(6 \;I,II)m=0}_{\frac{1}{2}} ={\cal N}_0  \; f^{-F_{56}\,- \tilde{F}_{56}+1/2} 
\stackrel{56}{(+)}\psi^{(4 \; I,II)}_{(+)}.
\label{Massless}
\end{eqnarray} 
The solutions~(Eq.\ref{Massless}) are the eigen functions  of $M^{56}$ with the eigen value $1/2$. 
Since no right handed massless solutions are allowed, the left handed ones are mass protected. 
For the  particular choice  $2(F_{56} + \tilde{F}_{56})=1$ the spin connection fields  
$-S^{56} \omega_{56\sigma} - \tilde{S}^{56} \tilde{\omega}_{56\sigma}$ 
compensate the term $\frac{1}{2Ef} \{p_{\sigma}, Ef \}_- $ and the  left handed spinor
with respect to $d=(1+3)$ becomes a constant with respect to $\rho $ and $\phi$.  
To make one of these two states massive, one can try to include terms like 
$\tilde{F}^{\cmp i}$.

Let us keep $\tilde{F}^{\oplus i}=0\; i\in\{1,2,3\}$ and $F_{56}= \tilde{F}_{56}$, while we take 
$\tilde{F}^{\oplus 3}\,,\tilde{F}^{\oplus \smp}$ non zero.
Now it is still true that due to the conditions in Eq.~(\ref{norm-solAB1234})  there are no 
massless solutions determined by ${\cal A}^{III,VI}$ and ${\cal B}^{III,VI}$. 
There is now only one massless and mass protected family for $F_{56}= \tilde{F}_{56}$. In this case the 
solutions $\mathcal{A}^{I -}_{0} $ and $\mathcal{A}^{II -}_{0} $ are related
\begin{eqnarray}
\label{masslessAm}
\mathcal{A}^{I -}_{0} &=& {\cal N}^{\pm}_{0} \, f^{\frac{1}{2}[1-2F_{56}-2\tilde{F}_{56}-
      2 \sqrt{(\tilde{F}^{\ominus 3})^2+\tilde{F}^{\ominus \boxplus} \tilde{F}^{\ominus \boxminus}}]} 
      \,,\nonumber\\ 
\mathcal{A}^{II -}_{0} &=& - \frac{(\sqrt{(\tilde{F}^{\ominus 3})^2+ \tilde{F}^{\ominus \boxplus}
\tilde{F}^{\ominus \boxminus} }+ \tilde{F}^{\ominus 3})}{\tilde{F}^{\ominus \boxplus}}\,\mathcal{A}^{I -}_{0}\,.
\end{eqnarray}
There exists, however, one additional massless state, with $\mathcal{A}^{I +}_{0} $  
related to $\mathcal{A}^{II + }_{0} $ and $\mathcal{B}^{I +}_{0} $ 
related to $\mathcal{B}^{II + }_{0} $, which fulfil Eq.~(\ref{norm-solAB1234}). But since we have left 
and right handed massless solution present, it is not mass protected any longer.

One can make  a choice as well that none of solutions would be massless.

Let us conclude this section by recognizing that for $\tilde{F}^{\oplus \spm}=0 $ and  
$\tilde{F}^{\ominus \spm} =0$ all the families decouple. There is then the choice of the parameters 
($F_{56}\,$, $\tilde{F}_{56}\,$, $\tilde{F}^{\oplus 3}$, $\tilde{F}^{\ominus 3}\, $) which determine 
how many massless and mass protected families exist, if  any.

\section{Conclusions and discussions}
\label{conclusion}

We make in this contribution a small step further with respect to the ref.~\cite{dhn} in  
understanding the existence of massless and mass protected spinors in non compact spaces 
in the presence of families of spinors after breaking symmetries. 
We take  a toy model in ${\cal M}^{5+1}$, which breaks into ${\cal M}^{3+1} \times$  an 
infinite disc curled into an almost $S^2$ under the influence of the zweibein. Following the 
{\it spin-charge-family} theory we have in this toy model four families. We study properties of 
families when allowing that besides  
the spin connection field  - the gauge field of $S^{st}=\frac{i}{4}(\gamma^s \gamma^t-
\gamma^t \gamma^s)$ - also the gauge fields of $\tilde{S}^{st}=\frac{i}{4}
(\tilde{\gamma}^s \tilde{\gamma}^t- \tilde{\gamma}^t \tilde{\gamma}^s)$ determine families 
properties (as suggested by the {\it spin-charge-family} theory).   

We simplify our study by assuming the same radial dependence of all the spin connection fields, 
the one used in studies without families, allowing here that only  strengths of the field vary 
within some intervals. We take these strengths as parameters, which are allowed to change.

We found that the choices of the parameters allows within some intervals four, two or 
none massless  and mass protected spinors.

Families like obviously to be massless  and mass protected in even numbers.

\end{document}